\documentclass[superscriptaddress,twocolumn,showpacs]{revtex4-2}
\usepackage[dvipdfmx]{graphicx}
\DeclareUnicodeCharacter{FF0C}{,}
\usepackage{color}
\usepackage{dcolumn}
\usepackage{amsmath}
\usepackage{lipsum}
\usepackage{dsfont}
\usepackage{epsfig}
\usepackage{soul}
\usepackage{bm}
\usepackage{amssymb}
\usepackage[dvipsnames]{xcolor}
\newcommand{\be}{\begin{equation}}
\newcommand{\ee}{\end{equation}}

\begin{document}
\title{Opening and closing a bandgap via alternating softening and hardening nonlinearities}

\author{Weijian Jiao}
\email{wjiao@tongji.edu.cn}
\affiliation{School of Aerospace Engineering and Applied Mechanics, Tongji University, Shanghai 200092, China}
\affiliation{Shanghai Institute of Aircraft Mechanics and Control,  Shanghai, 200092, China}

\begin{abstract}
Recent studies have shown some unusual nonlinear dispersion behaviors that are disconnected from the linear regime. However,  existing analytical techniques, such as perturbation methods, fail to correctly capture these behaviors. Here we propose a general theoretical approach that converts the nonlinear wave equation to an equivalent linear eigenvalue problem, which directly gives the nonlinear dispersion relation and modal vectors. The theoretical approach is employed to 1D phononic chains and 2D hexagonal lattices with alternating softening and hardening nonlinearities, revealing amplitude-induced bandgap opening and closing phenomena. The theoretical results are validated via full-scale simulations with periodic boundary conditions， in which steady-state nonlinear plane wave responses are numerically obtained. Moreover, we leverage these nonlinear phenomena to achieve tunable frequency splitting and focusing effects. Thus, our work opens new paradigms for understanding nonlinear wave physics and for achieving novel wave control capabilities.
\end{abstract}

\maketitle
\section{Introduction}

Phononic crystals and metamaterials have been known for their unusual abilities to manipulate wave propagation \cite{Hussein_2014}. An important ability is to prohibit wave propagation in certain frequency intervals, which are referred to as bandgaps. There are two main categories of bandgaps -- Bragg bandgaps resulting from periodicity \cite{brillouin1946wave} and locally resonant bandgaps enabled by local resonant mechanisms \cite{liu2000locally}. For practical applications, bandgaps have been extensively exploited for sound attenuation and vibration control \cite{martinez1995sound, Celli2015,jin2021physics}. In addition to bandgaps, researchers have realized a variety of unique wave properties, including negative refraction \cite{Zhu_NC_2014}, topological modes \cite{Kane_NP_2014,pal2017edge,Ma_PRL_2018,huang2024parity}, and nonreciprocal wave propagation \cite{bunyan2018acoustic,Wang_PRL_2018,celli2024time,yuan2024nonreciprocal}. 

Recently, significant efforts have been devoted to achieving tunable properties \cite{bertoldi2017flexible,Kochmann_review_2017,wang2020tunable}. Here, we recall some notable strategies, such as instabilities \cite{Wang_PRL_2014,Deng_PNAS_2020}, electro-elastic coupling \cite{celli2015tunable, wu2018tuning}, magneto-elastic coupling \cite{Bilal_2017,Kuan_PRB_2021,zhang2025topological}, and phase-transforming \cite{Liang_pnas_2022,zou2023magneto,jiao2024toward,stenseng2025bi}. Another interesting strategy is to leverage the amplitude-dependent characteristics of nonlinear waves \cite{boechler2011tunable,fang2017ultra,patil2022review,fronk2023_ND,bordiga2024automated}. For weakly nonlinear harmonic waves, tunability is enabled by amplitude-dependent nonlinear dispersion relation, which can manifest either as frequency shift or wavenumber shift depending
on whether the excitation is imposed as an initial
condition or as a boundary condition, respectively \cite{Jiao_prsa_2021}. Different techniques have been developed to analytically or numerically capture the nonlinear dispersion relation for various phononic systems, among which the popular ones are the transfer matrix
method \cite{manktelow2013comparison,khajehtourian2014dispersion,zhao2024wideband}, harmonic balance methods (HBM) \cite{lazarov2007low,narisetti2012study,abedinnasab2013wave,yi2024}, and perturbation methods \cite{Narisetti_2011,zhou2018spectro,Jiao_pre_2019,fang2022perturbation}. 

In most studies, the effects of weak nonlinearity on dispersion relation are found to be small corrections to its linear counterpart. Interestingly, a few recent papers present some novel nonlinear dispersion properties that significantly deviate from linear properties \cite{yi2024,fronk2023computational,bae2025opening,sone2024nonlinearity}. For example, Bae et al. \cite{bae2025opening}  proposed a monoatomic-diatomic convertible metamaterial with alternating nonlinearity to open a bandgap, which is theoretically revealed by applying \textit{Brillouin-Wigner} perturbation instead of the general \textit{Lindstedt-Poincaré} perturbation. Yi and Chen \cite{yi2024} employed the Galerkin method and HBM to numerically calculate the nonlinear dispersion relation of a high-order topological kagome lattice with intercell hardening and intracell softening nonlinearities, showing an unusual bandgap crossing effect. For topological systems, researchers investigated nonlinearity-induced topological phase transitions, in which the existence of topological edge modes depends on the mode amplitude and they are not available in the linear regime \cite{zhou2022topological,sone2024nonlinearity}.

While previous studies have shown some unusual dispersion behaviors enabled by alternating nonlinearity, a general and robust theoretical approach has been lacking, especially but not only in the context of phononic crystals and metamaterials. In addition, full-scale numerical simulations that yield steady-state plane wave responses have rarely been performed to directly validate these nonlinear dispersion behaviors. In this work, we propose a general theoretical approach that converts the nonlinear wave equation to an equivalent linear eigenvalue problem, which directly gives the nonlinear dispersion relation and the nonlinear modal vectors. By applying the presented approach to 1D nonlinear mass-spring chains and 2D nonlinear hexagonal lattices, we theoretically reveal amplitude-induced bandgap opening and closing phenomena enabled by alternating softening and hardening nonlinearities in the springs. Moreover, we numerically obtain steady-state plane wave responses via full-scale simulation with periodic boundary conditions, unequivocally confirming the theoretically predicted nonlinear dispersion and modal behaviors. 

This work provides a general theoretical approach to solve nonlinear wave problems, which overcomes the limitations of regular perturbation methods for phononic systems with alternating nonlinearities. Furthermore, the bandgap opening and closing phenomena enable nonlinear systems with superior tunability, which can open a new paradigm for wave control study.      

\section{Wave propagation in 1D nonlinear mass-spring chains}
\begin{figure}[htbp]
    \centerline{ \includegraphics[width=0.5\textwidth]{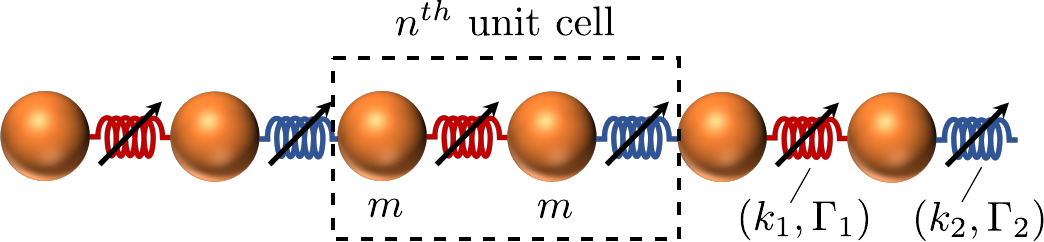}}
    \caption{Schematic of a one-dimensional chain of identical masses connected by alternating two different nonlinear springs.} 
    \label{fig:Schematic_monatomic}
\end{figure}
\subsection{Theoretical approach}\label{Theory}

\begin{figure*}[htbp]
    \centerline{ \includegraphics[width=0.9\textwidth]{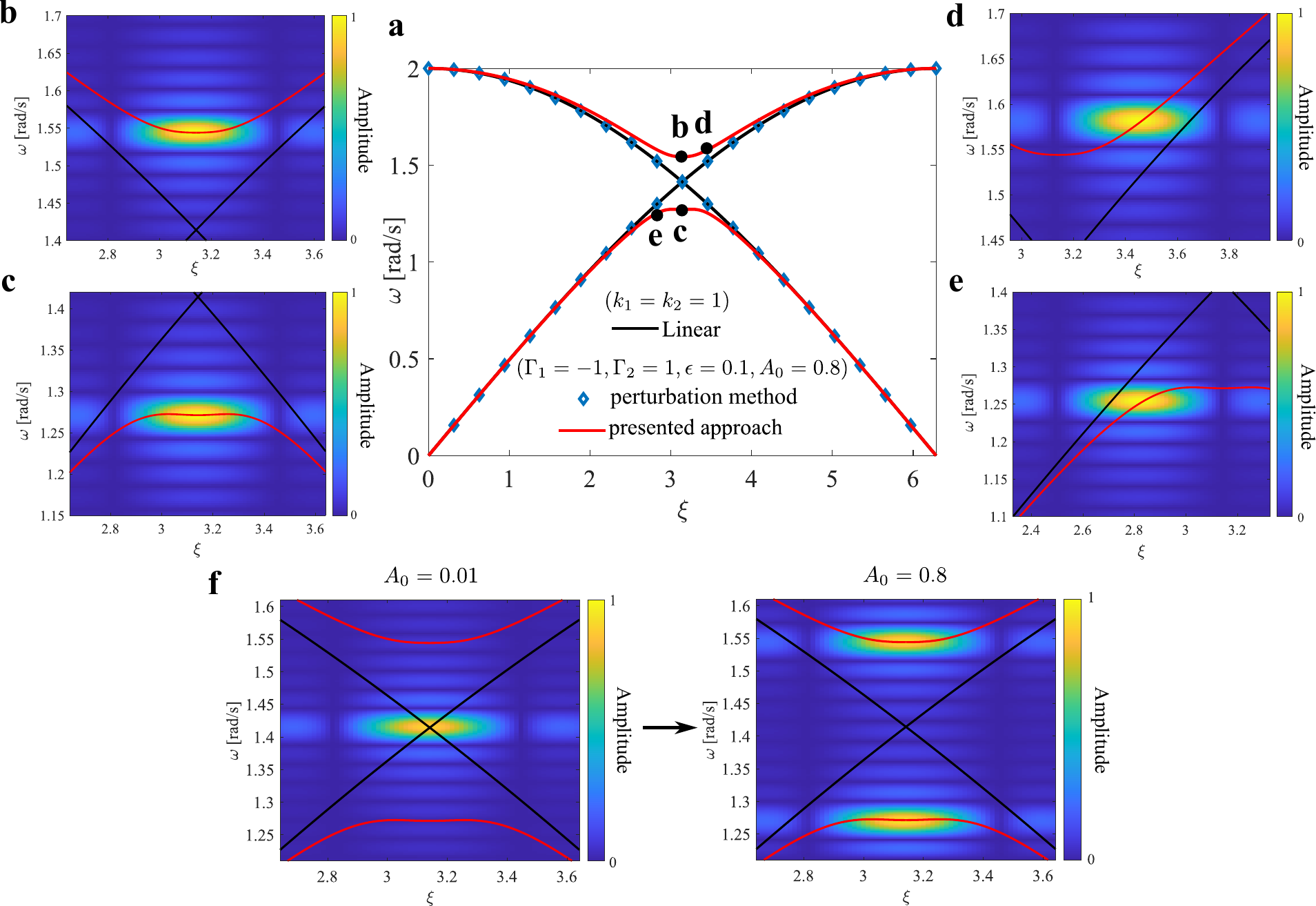}}
    \caption{Bandgap opening in a nonlinear monatomic-like chain. \textbf{a} Dispersion relations of a monatomic-like chain with parameters $m=1$, $k_1=k_2=1$,  $\epsilon=0.1$, $\Gamma_1=-1$ and $\Gamma_2=1$. The linear dispersion relation is plotted as black curves, with no bandgap. For excitation amplitude $A_0=0.8$, a nonlinear dispersion relation is obtained from the presented approach  (red curves), displaying a bandgap formed at $\xi=\pi$. A nonlinear dispersion relation is obtained from a general perturbation method (diamond markers), identical to the linear one. \textbf{b-e} Numerically-obtained spectra for four dispersion points highlighted by black dots in panel \textbf{a}.  \textbf{f} Frequency splitting effect achieved by increasing the excitation amplitude from $A_0=0.01$ to $A_0=0.8$.} 
    \label{fig:Bandgap_opening}
\end{figure*}

\begin{figure*}[htbp]
    \centerline{ \includegraphics[width=0.9\textwidth]{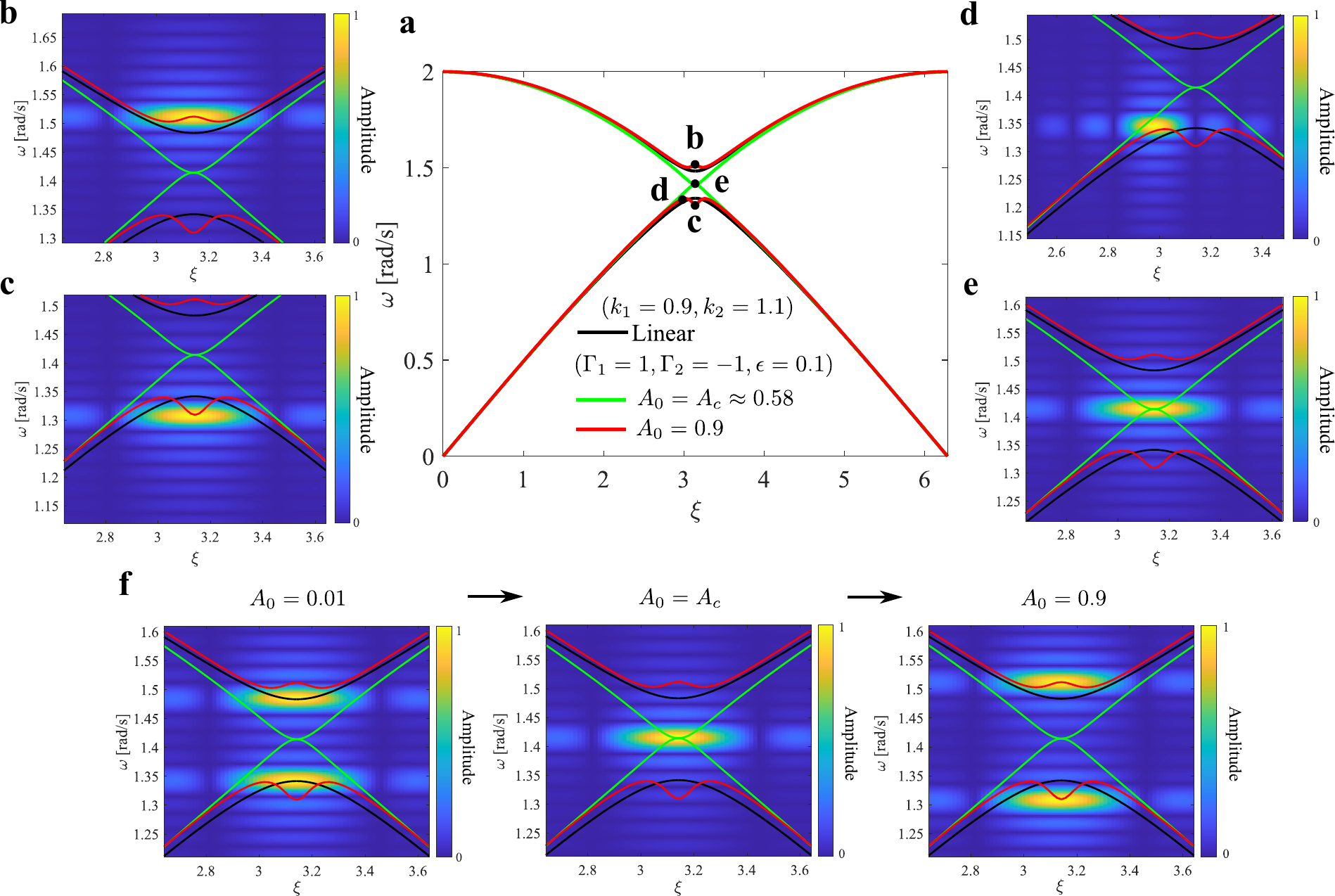}}
    \caption{Bandgap closing and reopening in a nonlinear diatomic-like chain. \textbf{a} Dispersion relations of a diatomic-like chain with parameters $m=1$, $k_1=0.9$, $k_2=1.1$,  $\epsilon=0.1$, $\Gamma_1=1$ and $\Gamma_2=-1$: linear dispersion relation (black curves), displaying a Bragg bandgap; nonlinear dispersion relation  (red curves) obtained for the critical amplitude $A_0=A_c\approx0.58$, with no bandgap; nonlinear dispersion relation  (green curves) obtained for excitation amplitude $A_0=0.9$, displaying a bandgap. \textbf{b-e} Numerically-obtained spectra for four dispersion points highlighted by black dots in panel \textbf{a}.  \textbf{f} Frequency focusing and re-splitting effect achieved by increasing the excitation amplitude from $A_0=0.01$ to $A_0=A_c$ to $A_0=0.9$.} 
    \label{fig:Bandgap_closing}
\end{figure*}

We first consider a mass-spring chain of identical masses $m$ connected by alternating two different nonlinear springs (Fig.~\ref{fig:Schematic_monatomic}), which exhibit weakly cubic nonlinear force-displacement behavior, i.e., $f_{1,2}=k_{1,2}\delta+\epsilon\Gamma_{1,2}\delta^3$ ($k_{1,2}$ and $\epsilon\Gamma_{1,2}$ are the linear and nonlinear spring constants, and $\epsilon$ is a small parameter). The equations of motion for the $n^{th}$ unit cell can be written as
\begin{align}\label{EOMs_monatomic}
\begin{split}
m\ddot{u}^a_{n}&=k_1(u^b_{n}-u^a_{n})+\epsilon\Gamma_1(u^b_{n}-u^a_{n})^3\\
&+k_2(u^b_{n-1}-u^a_{n})+\epsilon\Gamma_2(u^b_{n-1}-u^a_{n})^3,\\
m\ddot{u}^b_{n}&=k_2(u^a_{n+1}-u^b_{n})+\epsilon\Gamma_2(u^a_{n+1}-u^b_{n})^3\\
&+k_1(u^a_{n}-u^b_{n})+\epsilon\Gamma_1(u^a_{n}-u^b_{n})^3,
\end{split}
\end{align}
where $u^{a,b}$ denote the displacements of the two masses in the unit cell.

Here, we propose a new approach to investigate the propagation of nonlinear plane wave in the mass-spring system, which is assumed to take the form with the Bloch ansatz $e^{i\theta_n}$:
\begin{equation}\label{Harmonic_soln}
\bold{u}_{n}=\begin{bmatrix} u^a_n \\ u^b_n \end{bmatrix}=
 \frac{A}{2}\boldsymbol{\psi}e^{i\theta_n}+\frac{A}{2}\boldsymbol{\psi}^*e^{-i\theta_n},
\end{equation}
where $A$ denotes wave amplitude, $\boldsymbol{\psi}=[\psi^a\,\,\,  \psi^b]^T$ is a modal vector, $\theta_n=\xi n-\omega t$ is a spatiotemporal variable where $\xi$ and $\omega$ are the nondimensional wavenumber and the frequency, respectively,  and $(\cdot)^*$ denotes the complex conjugate of a quantity. It should be noted that, for initial spatial excitation with prescribed wavenumber $\xi_0$,  ($\omega, \boldsymbol{\psi}$) are amplitude-dependent quantities to be determined, which can be different from the linear ones ($\omega_0, \boldsymbol{\psi}_0$).

The key idea of the presented approach is to convert the nonlinear problem to an equivalent linear eigenvalue problem. To this end, we first rewrite Eq.~\ref{EOMs_monatomic}, using the concept of equivalent dynamic stiffness, as
\begin{align}\label{EOMs_chains}
\begin{split}
m\ddot{u}^a_{n}&=k_1^{e}(u^b_{n}-u^a_{n})+k_2^{e}(u^b_{n-1}-u^a_{n}),\\
m\ddot{u}^b_{n}&=\Bar{k}_2^{e}(u^a_{n+1}-u^b_{n})+k_1^{e}(u^a_{n}-u^b_{n}),
\end{split}
\end{align}
where $k_1^{e}=k_1+\epsilon\Gamma_1(u^b_{n}-u^a_{n})^2$, $k_2^{e}=k_2+\epsilon\Gamma_2(u^b_{n-1}-u^a_{n})^2$, and $\Bar{k}_2^{e}=k_2+\epsilon\Gamma_2(u^a_{n+1}-u^b_{n})^2$  are defined as the equivalent dynamic stiffnesses. As an initial guess, we evaluate these equivalent siffinesses using the solution $\bold{u}_{n}(\theta_n, \boldsymbol{\psi}_0)$ with the linear eigenvector $\boldsymbol{\psi}_0$ obtained from Bloch analysis. To derive the equivalent linear eigenvalue problem, we keep only the terms associated with $e^{i\theta_n}$. Among these terms, we find that some contain the complex conjugate of the nonlinear eigenvector component, i.e., $(\psi^{a,b})^*$. To resolve this issue, we replace $(\psi^{a,b})^*$ by its linear counterpart $(\psi_0^{a,b})^*$ and return a linear eigenvector component $\psi_0^{a,b}$ back to the nonlinear one $\psi^{a,b}$. Finally, the nonlinear problem is successfully converted to the following linear eigenvalue problem:
\begin{equation}
\left[-\omega^2\bold{M}+\bold{K}(\xi_0, A,\boldsymbol{\psi}_0)\right]\boldsymbol{\psi}=\bold{0},\label{EOMs_v1}
\end{equation}
where $\bold{M}=\begin{bmatrix} m & 0\\ 0 & m \end{bmatrix}$ is a mass matrix and $\bold{K}(\xi_0, A,\boldsymbol{\psi}_0)=\begin{bmatrix} K_{11} & K_{12}\\ K_{21} & K_{22} \end{bmatrix}$ is a stiffness matrix depending on the imposed wavenumber $\xi_0$, wave amplitude $A$ and modal vector $\boldsymbol{\psi}_0$. The derivation detail is provided in Method, and the full expression of $\bold{K}$ is reported in SI (Note 1). 

For initial excitation with prescribed wavenumber $\xi_0$ and amplitude $A_0$, we can now obtain the nonlinear dispersion relation $(\xi_0,\omega)$ and modal vector $\boldsymbol{\psi}$ by solving Eq.\ref{EOMs_v1}. Once a nonlinear modal vector $\boldsymbol{\psi}_1$ is obtained, Eq.\ref{EOMs_v1} can be updated using $\bold{K}(\xi_0, A_0,\boldsymbol{\psi}_1)$, which gives rise to a corrected dispersion relation. Moreover, we can set a tolerance and repeat the above procedures until the relative residual of the modal vector $\left\Vert\Delta \boldsymbol{\psi}\right\Vert$ satisfies the tolerance. While this is reminiscent of iterative methods in computational mathematics, Eq.\ref{EOMs_v1} with updated $\bold{K}(\xi_0,A_0, \boldsymbol{\psi}_i)$ can be analytically solved at each step (see Method for details).

Compared to other existing methods in the literature (e.g., \textit{Lindstedt-Poincaré} method), the presented approach allows the modal vector $\boldsymbol{\psi}$ to be a variable. In the following, we will show that this is critical for theoretical discovery of the nonlinearity-induced phenomena of bandgap opening and closing that cannot be revealed using regular perturbation methods. 

\subsection{Bandgap opening in nonlinear monatomic-like chains}

Using the presented approach, we demonstrate how to open a bandgap in monatomic-like chains with alternating softening and hardening nonlinearities. To this end, we consider a nonlinear chain with the following parameters: $m=1$, $k_1=k_2=1$,  $\epsilon=0.1$, $\Gamma_1=-1$ and $\Gamma_2=1$ (standard SI units are adopted throughout the paper and omitted for simplicity). At low wave amplitude ($A\ll 1$), the system behaves like a monatomic chain, as the nonlinear effect can be ignored resulting in nearly identical equivalent dynamic stiffness $k^e_1 \approx k_1=k_2 \approx k^e_2$. At high wave amplitude, the system behaves like a diatomic chain, as the softening and hardening nonlinear effects are non-negligible, resulting in distinct equivalent dynamic stiffnesses $k^e_1 \neq k^e_2$. Thus, the system exhibits a bandgap at high amplitude, which vanishes as $A\ll 1$.

In Fig.~\ref{fig:Bandgap_opening}\textbf{a}, we show the linear (black curves) and nonlinear (red curves) dispersion relations, the latter of which is obtained for excitation amplitude $A_0=0.8$. Clearly, a newly-opened bandgap is captured by the nonlinear dispersion relation calculated from the presented approach (the tolerance is set to $10^{-6}$ for all cases in this study). For comparison, we plot the nonlinear dispersion relation (diamond markers) obtained from a multiple scales analysis, which is a perturbation method widely adopted in relevant studies \cite{holmes2012introduction,Ganesh_2017,Jiao_prsa_2021,zhou2022topological} (see Note 2 in SI for derivation). Surprisingly, these markers are well aligned with the linear dispersion relation with no bandgap opened, which indicates the inability of the perturbation analysis to capture the bandgap opening phenomenon. This inability results from the treatment that the displacement (and therefore the modal vector) is assumed as a small perturbation to the linear solution, which is obviously incorrect at wavenumbers near $\xi=\pi$ as the system is transformed from a monatomic chain to a diatomic chain due to nonlinear effects. This is also the case when using the general \textit{Lindstedt-Poincaré} perturbation method, as reported in \cite{bae2025opening}. 

To validate the above theoretical findings, we perform full-scale numerical simulations using the fourth-order Runge-Kutta method. Specifically, we consider a chain consisting of 40 identical masses connected by springs with alternating softening and hardening nonlinearities. To numerically establish a plane wave response with wavenumber $\xi_0$ and amplitude $A_0$, we apply initial spatial and velocity profiles over the entire chain using Eq.~\ref{Harmonic_soln} with the corresponding theoretically-obtained quantities $(\omega, \boldsymbol{\psi})$. Moreover, we impose the periodic boundary condition to the first and last masses. Then, the spatiotemporal response of the chain over a given time interval is recorded and transformed to the Fourier domain via two-dimensional discrete Fourier transform (2D-DFT). 

In the simulations, we consider four different dispersion points highlighted by black dots in Fig.~\ref{fig:Bandgap_opening}\textbf{a}. We observe that nearly steady-state plane wave responses are numerically achieved (see SI Figure and Movie.~1).
In Figs.~\ref{fig:Bandgap_opening}\textbf{b-e}, we show the corresponding spectra, matching well with the nonlinear dispersion relation. These results unequivocally confirm the bandgap opening phenomenon in the monatomic-like chain and the validity of the presented theoretical approach. 

The bandgap opening phenomenon enables the system with superior wave control capabilities. In addition to bandgap tuning, we demonstrate the effect of frequency splitting by simply controlling the excitation amplitude $A_0$. Specifically, we prescribe the initial spatial profile as $\bold{u}^0_n=A_0(\boldsymbol{\psi}^1_{\pi}+\boldsymbol{\psi}^2_{\pi})$, where $\boldsymbol{\psi}^1_{\pi}$ and $\boldsymbol{\psi}^2_{\pi}$ are the nonlinear modal vectors obtained at $\xi_0=\pi$ for the acoustic and optical branches, respectively. As shown in Fig.~\ref{fig:Bandgap_opening}\textbf{e} (Movie.~2), the response is well aligned with the linear dispersion relation and characterized by a single frequency component for low excitation amplitude $A_0=0.01$. In contrast, it splits into two frequency components at high excitation amplitude $A_0=0.8$, as the initial profile is a superposition of the two nonlinear modal shapes.

\subsection{Bandgap closing and reopening in nonlinear diatomic-like chains}

Next, we demonstrate bandgap closing and reopening phenomena in nonlinear diatomic-like chains with the following parameters: $m=1$, $k_1=0.9$, $k_2=1.1$,  $\epsilon=0.1$, $\Gamma_1=1$ and $\Gamma_2=-1$. At low amplitude $A\ll1$, the system behaves like a diatomic chain with a bandgap, as the nonlinear effect can be ignored, resulting in two different equivalent dynamic stiffnesses $k^e_1\approx k_1 < k_2\approx k^e_2$. As wave amplitude increases, $k^e_1$ increases while $k^e_2$ decreases, so it is expected that the bandgap can be narrowed down. In principle, there should exist a critical amplitude $A_c$, leading to the case where the bandgap is just closed (see Method for the determination of $A_c$). It is worth noting that $A_c$ needs to be determined from Eq.~\ref{EOMs_v1}, which does not necessarily correspond to the condition $k^e_1=k^e_2$. As wave amplitude continues to increase, the system may behave like a diatomic chain again for $k^e_1>k^e_2$ cases, leading to the bandgap reopening phenomenon.

In Fig.~\ref{fig:Bandgap_closing}\textbf{a}, we plot the linear dispersion relation (black curves) and the nonlinear dispersion relations obtained from the presented approach for excitation amplitudes $A_0=A_c\approx0.58$ (green curves) and $A_0=0.9$ (red curves). Here, we theoretically show that the system has a bandgap at low amplitudes, which is closed at the critical amplitude $A_c$ and reopened at high amplitude $A_0=0.9$. To confirm these findings, we follow the procedures described above to perform numerical simulation and spectral analysis for four dispersion points indicated by black dots in Fig.~\ref{fig:Bandgap_closing}\textbf{a}. To resolve the wavenumber $\xi_0=0.95\pi$ corresponding to point \textbf{d}, the chain is simulated with 80 masses, instead of 40 masses for the other three points. In Fig.~\ref{fig:Bandgap_closing}\textbf{b-e}, we show the spectra for the four dispersion points, which are obtained from the nonlinear plane wave responses (see Movie.~3). We show that the numerical results are in excellent agreement with the theoretical predictions, thereby confirming the phenomenon of bandgap closing and reopening in the nonlinear diatomic-like chain.

Moreover, we leverage this phenomenon to achieve frequency focusing and re-splitting effect via a simple control of the excitation amplitude $A_0$. Similar to the previous treatment, we numerically demonstrate this effect by prescribing the initial spatial profile as a superposition of the two modal vectors obtained at $\xi_0=\pi$, i.e., $\bold{u}^0_n=A_0(\boldsymbol{\psi}^1_{\pi}+\boldsymbol{\psi}^2_{\pi})$. As shown in Fig.~\ref{fig:Bandgap_closing}\textbf{f}, the response has two frequency components for low excitation amplitude $A_0=0.01$. As the amplitude increases to the critical amplitude $A_0=A_c$, the response is characterized by a single frequency component, as a result of the bandgap closing phenomenon. When the amplitude reaches $A_0=0.9$, the spectral response splits into two components as the bandgap is reopened (Movie.~4). 

\section{Wave propagation in nonlinear hexagonal lattices}

\begin{figure}[htbp]
    \centerline{ \includegraphics[width=0.5\textwidth]{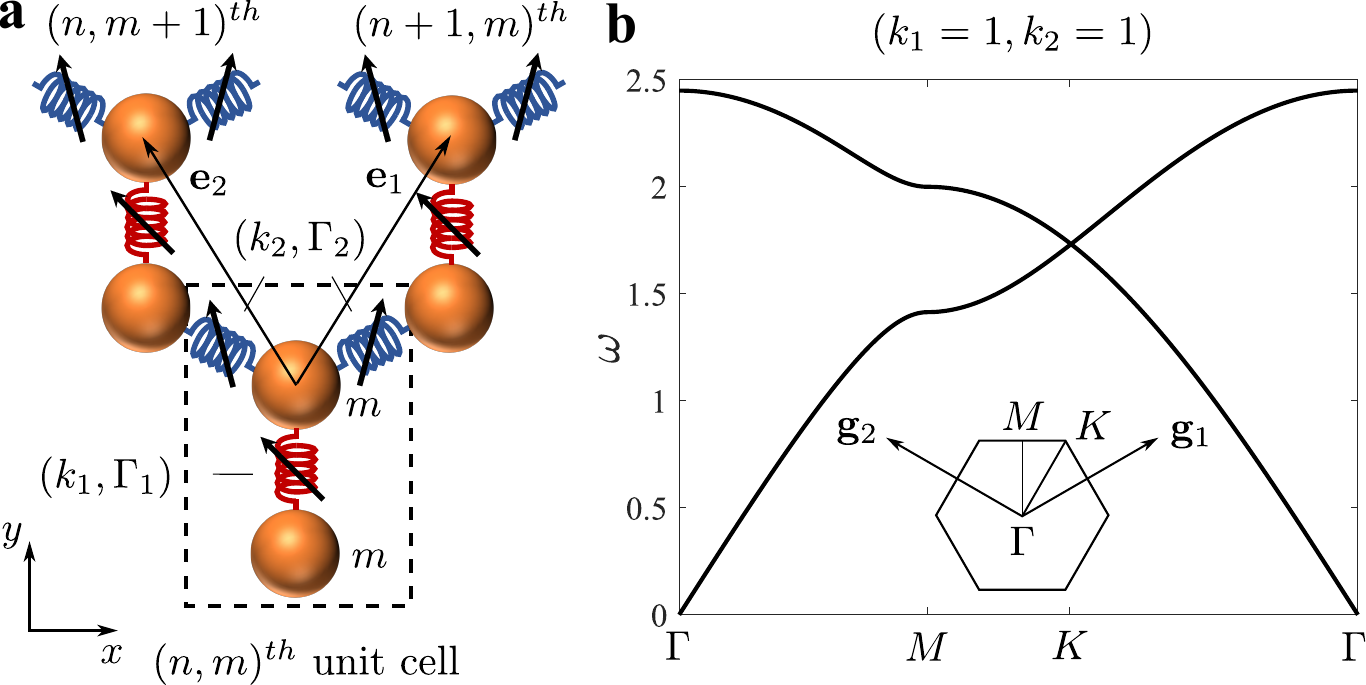}}
    \caption{\textbf{a} Schematic of a hexagonal lattice with different nonlinearities. \textbf{b} Linear dispersion relation along the contour of the irreducible Brillouin zone, showing a Dirac cone at the $K$ point.} 
    \label{fig:Schematic_hexagonal}
\end{figure}

\begin{figure*}[htbp]
    \centerline{ \includegraphics[width=0.85\textwidth]{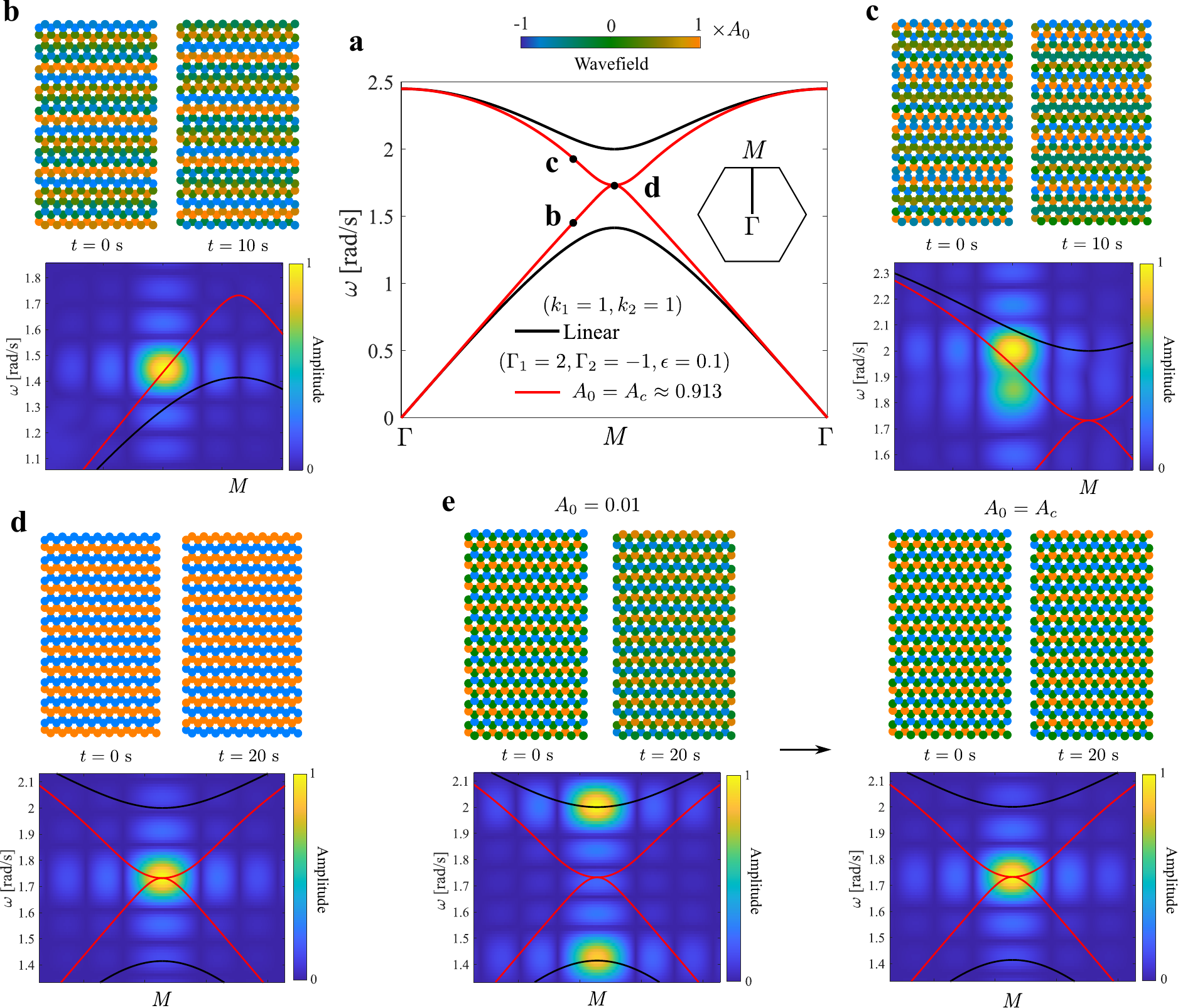}}
    \caption{Bandgap closing in a nonlinear hexagonal lattice. \textbf{a} Dispersion relations along the $\Gamma-M$ direction of a hexagonal lattice with parameters $m=1$, $k_1=1$, $k_2=1$,  $\epsilon=0.1$, $\Gamma_1=2$ and $\Gamma_2=-1$: linear dispersion relation (black curves), displaying a bandgap; nonlinear dispersion relation  (red curves) obtained for the critical amplitude $A_0=A_c\approx0.913$, with no bandgap. \textbf{b-e} Numerically-obtained wavefields (top) and spectra (bottom) for three dispersion points highlighted by black dots in panel \textbf{a}.  \textbf{d} Frequency focusing effect achieved by increasing the excitation amplitude from $A_0=0.01$ to $A_0=A_c$.} 
    \label{fig:Hexagonal_M}
\end{figure*}

\begin{figure*}[htbp]
    \centerline{ \includegraphics[width=0.9\textwidth]{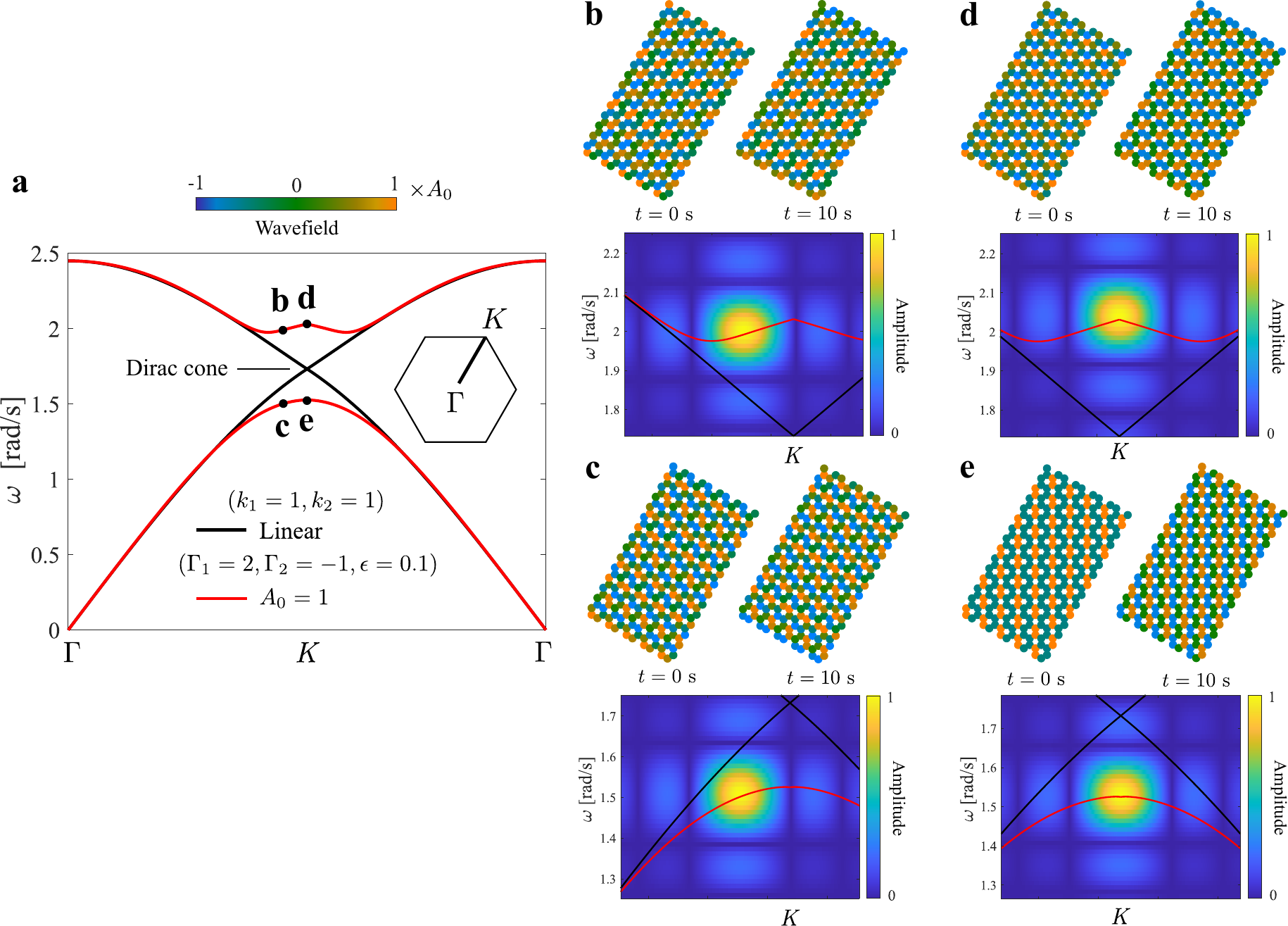}}
    \caption{Bandgap opening in a nonlinear hexagonal lattice. \textbf{a} Dispersion relations along the $\Gamma-K$ direction of a hexagonal lattice with parameters $m=1$, $k_1=1$, $k_2=1$,  $\epsilon=0.1$, $\Gamma_1=2$ and $\Gamma_2=-1$: linear dispersion relation (black curves), with no bandgap; nonlinear dispersion relation  (red curves) obtained for excitation amplitude $A_0=1$, displaying a bandgap formed at the $K$ point. \textbf{b-e} Numerically-obtained wavefields (top) and spectra (bottom) for four dispersion points highlighted by black dots in panel \textbf{a}.} 
    \label{fig:Hexagonal_K}
\end{figure*}

In this section, we extend the above analysis to 2D nonlinear hexagonal lattices. As shown in Fig.~\ref{fig:Schematic_hexagonal}\textbf{a}, we consider a hexagonal lattice with lattice vectors $\bold{e}_1=L[\cos(\pi/3) \,\, \sin(\pi/3)]^T$ and $\bold{e}_2=L[-\cos(\pi/3) \,\, \sin(\pi/3)]^T$, where $L$ is the distance between next-nearest neighbors.  The unit cell contains two identical masses $m$ connected by a spring exhibiting a cubic nonlinear force-displacement relation $f_1=k_1\delta+\epsilon\Gamma_1\delta^3$, while the springs connecting neighboring unit cells have a different set of spring constants $(k_2,\Gamma_2)$. The equations of motion for the $(n,m)^{th}$ unit cell at position $\bold{r}_{n,m}=n\bold{e}_1+m\bold{e}_2$ can be written as 

\begin{align}\label{EOMs_hexagonal}
\begin{split}
m\ddot{w}^a_{n,m}&=k_1^{e}(w^b_{n,m}-w^a_{n,m})+k_2^{e}(w^b_{n-1,m}-w^a_{n,m})\\
&+k_3^{e}(w^b_{n,m-1}-w^a_{n,m}),\\
m\ddot{w}^b_{n,m}&=k_1^{e}(w^a_{n,m}-w^b_{n,m})+\bar{k}_2^{e}(w^a_{n+1,m}-w^b_{n,m})\\
&+\bar{k}_3^{e}(w^a_{n,m+1}-w^b_{n,m}),\\
\end{split}
\end{align}
where $k_1^{e}=k_1+\epsilon\Gamma_1(w^b_{n,m}-w^a_{n,m})^2$, $k_2^{e}=k_2+\epsilon\Gamma_2(w^b_{n-1,m}-w^a_{n,m})^2$, 
$k_3^{e}=k_2+\epsilon\Gamma_2(w^b_{n,m-1}-w^a_{n,m})^2$, $\Bar{k}_2^{e}=k_2+\epsilon\Gamma_2(w^a_{n+1,m}-w^b_{n,m})^2$, and $\Bar{k}_3^{e}=k_2+\epsilon\Gamma_2(w^a_{n,m+1}-w^b_{n,m})^2$ are the equivalent dynamic stiffnesses.

Following the same theoretical treatment discussed in Method for 1D phononic chains, Eq.~\ref{EOMs_hexagonal} can be converted to the following linear eigenvalue problem
\begin{equation}
\left[-\omega^2\bold{M}+\bold{K}(\boldsymbol{\kappa}_0, A,\boldsymbol{\psi}_{i-1})\right]\boldsymbol{\psi}_i=\bold{0},\label{EOMs_hexagonal_eig}
\end{equation}
where $\bold{M}=\begin{bmatrix} m & 0\\ 0 & m \end{bmatrix}$ is a mass matrix, $\bold{K}(\boldsymbol{\kappa}_0, A,\boldsymbol{\psi}_{i-1})$ is a stiffness matrix (see Note 3 in SI for full expression), $\boldsymbol{\kappa}_0$ is the prescribed wave vector, $A$ is wave amplitude, and $\boldsymbol{\psi}_{i-1}$ and $\boldsymbol{\psi}_i$ are eigenvectors at the previous and current steps, respectively. Then, we can set a tolerance and begin to solve Eq.~\ref{EOMs_hexagonal_eig} using the linear solution as an initial guess. Once the tolerance is met, we obtain the nonlinear dispersion relation $(\xi_0,\omega)$ and the nonlinear modal shape $\boldsymbol{\psi}$. Accordingly, the nonlinear plane wave solution can be expressed as
\begin{equation}\label{Harmonic_soln_hexagonal}
\bold{w}_{n,m}=\begin{bmatrix} w^a_{n,m} \\ w^b_{n,m} \end{bmatrix}=
 \frac{A}{2}\boldsymbol{\psi}e^{i\theta_n}+\frac{A}{2}\boldsymbol{\psi}^*e^{-i\theta_n},
\end{equation}
where $\theta_n=\boldsymbol{\kappa}_0 \cdot \bold{r}_{n,m}- \omega t$ and $\boldsymbol{\kappa}_0$ can be expressed using the basis vectors of the reciprocal lattice $\bold{g}_1=4\pi[\cos(\pi/6) \,\, \sin(\pi/6)]^T/(\sqrt{3}L)$ and $\bold{g}_2=4\pi[-\cos(\pi/6) \,\, \sin(\pi/6)]^T/(\sqrt{3}L)$. 

In Fig.~\ref{fig:Schematic_hexagonal}\textbf{b}, we plot the linear dispersion relation along the contour of the irreducible Brillouin zone (IBZ) of a hexagonal lattice with parameters: $L=1$, $m=1$, $k_1=k_2=1$, $\epsilon=0.1$, $\Gamma_1=2$, and $\Gamma_2=-1$. Interestingly, the lattice exhibits rich direction-dependent wave characteristics. For example, there exists a bandgap for wave propagating along the $\Gamma-M$ direction, while there is no bandgap for wave propagating along the $\Gamma-K$ direction. 

In the following sections, we will show that the bandgap in the $\Gamma-M$ direction can be closed and a new bandgap can be opened in the $\Gamma-K$ direction. More interestingly, these phenomena are achieved within the same hexagonal lattice and through a simple control of the excitation amplitude.

\subsection{Bandgap closing in nonlinear hexagonal lattices along the $\Gamma-M$ direction}

Here, we first demonstrate the bandgap closing phenomenon in the $\Gamma-M$ direction. Using the same strategy described in Method, we determine the critical amplitude $A_c\approx 0.913$ and plot the corresponding nonlinear dispersion relation in Fig.~\ref{fig:Hexagonal_M}\textbf{a} (red curves), in which the bandgap is closed. 

Next, we perform a suite of full-scale simulations to corroborate the above findings. To this end, we consider a hexagonal lattice of $20\times 40$ masses along the $\Gamma-M$ direction. To establish plane wave responses, we prescribe the initial spatial profile using Eq.~\ref{Harmonic_soln_hexagonal} and impose the periodic boundary conditions in both the $x$ and $y$ directions. After the lattice is simulated for a sufficiently long time interval, we extract the spatiotemporal response from the masses located along the propagation direction (i.e, the $y$ direction), which is then transformed to the Fourier domain via 2D-DFT.

In Figs.~\ref{fig:Hexagonal_M}\textbf{b-d}, we display the wavefields and the corresponding spectra for three dispersion points highlighted by black dots in Fig.~\ref{fig:Hexagonal_M}\textbf{a}. From these wavefields, we observe that steady-state plane wave responses are numerically simulated over sufficiently long time periods for all cases (see Movie.~5). Moreover, the spectra match well with the nonlinear dispersion relation, which confirms the bandgap closing phenomenon for wave propagating along the $\Gamma-M$ direction with the critical amplitude $A_c$. Furthermore, we employ this phenomenon and a superpositioned initial spatial profile (i.e., $\bold{w}^0_{n,m}=A_0(\boldsymbol{\psi}^1_{\pi}+\boldsymbol{\psi}^2_{\pi})$ where $\boldsymbol{\psi}^1_{\pi}$ and $\boldsymbol{\psi}^2_{\pi}$ are the linear modal vectors obtained at $\boldsymbol{\kappa}_0=\bold{g}_1/2+\bold{g}_2/2$) to achieve amplitude-induced frequency focusing effect, as demonstrated in Fig.~\ref{fig:Hexagonal_M}\textbf{e} (Movie.~6).

\subsection{Bandgap opening in nonlinear hexagonal lattices along the $\Gamma-K$ direction}

In Fig.~\ref{fig:Hexagonal_K}\textbf{a}, we plot the nonlinear dispersion relation (red curves) in the $\Gamma-K$ direction for excitation amplitude $A_0=1$. For reference, we superimpose the linear dispersion relation indicated by black curves, in which a Dirac cone is observed at the $K$ point. As wave amplitude increases, the inversion symmetry of the lattice is broken by changing the parity of the springs, leading to the opening of a bandgap at the Dirac point.

To numerically confirm this phenomenon, we follow similar numerical procedures described in the previous section to conduct full-scale simulations. Here, we consider a hexagonal lattice of $20\times 40$ masses along the $\Gamma-K$ direction, and we apply the periodic boundary conditions in the $\Gamma-K$ direction and its orthogonal direction (i.e., $\bold{g}_2$). In Fig.~\ref{fig:Hexagonal_K}\textbf{b-e}, we show the wavefields and the corresponding spectra for four dispersion points highlighted by black dots in Fig.~\ref{fig:Hexagonal_K}\textbf{a}. From the numerical results, we observe nearly steady-state plane wave responses propagating along the $\Gamma-K$ direction (Movie.~7).  Moreover, the spectra are well aligned with the nonlinear dispersion relation, showing the formation of a bandgap at the $K$ point.   

Last, it is worth pointing out that the nonlinearity-induced bandgap is a partial bandgap, which only exists in the $\Gamma-K$ direction indicated in the inset of  Fig.~\ref{fig:Hexagonal_K}\textbf{a} and may not be valid in other $\Gamma-K$ directions. This can be ascribed to the fact that the six-fold symmetry of the lattice is broken as a result of breaking the parity of the springs via the softening and hardening nonlinear effects. 

\section{Discussions}

The fundamental assumption of the presented theoretical approach is the form of nonlinear plane wave solutions. It is therefore reasonable to claim that the approach works for different combinations of nonlinearities and even beyond the weakly nonlinear regime, as long as the nonlinear plane wave is physically obtainable. The existence of a nonlinear plane wave can be demonstrated through numerical simulation. However, the excitation condition plays a crucial role in establishing a steady-state nonlinear plane wave response, as the nonlinear effect manifests differently for different types of excitation \cite{Jiao_prsa_2021}. In general, it is more feasible (if possible) to obtain a nonlinear plane wave with initial condition than with boundary excitation.

As for future work, there are a few interesting things worth exploring.  For experimental demonstration, structural nonlinearity \cite{zhao2024wideband} and piezoelectric shunts \cite{alfahmi2024programmable} can be exploited to realize softening and hardening nonlinearities in mechanical systems.  In addition to nonlinear bulk modes, nonlinear topological modes have recently received significant attention due to their unique dynamic characteristics \cite{Bryan_pnas_2014,vila2019role,Zhou_PRB2020,zhou2022topological,xiu2023synthetically}. It may also be interesting to explore these topological modes with alternating nonlinearities. Finally, the idea of the theoretical approach to derive an equivalent linear eigenvalue problem for nonlinear plane waves in phononic systems may be extended to other physical systems. 

\section*{Acknowledgments}
This work was supported by the NSF of China (grant number 12402105), the Fundamental Research Funds for the Central Universities, and Shanghai Gaofeng Project for University Academic Program Development. W. J. sincerely appreciates the support from Lijuan Yu.

 \section*{Method}
\subsection{Derivation of the equivalent linear eigenvalue problem for 1D nonlinear chains}

We start by evaluating the equivalent dynamic stiffnesses in Eq.\ref{EOMs_chains} using the solution $\bold{u}_{n}(\theta_n, \boldsymbol{\psi}_0)$ with the linear eigenvector $\boldsymbol{\psi}_0$. Taking $k_1^{e}$ for example, we have 
\begin{widetext}
\begin{align}\label{stiffnesses_equiv}
\begin{split}
k_1^{e}&=k_1+\epsilon\Gamma_1(u^b_{n}-u^a_{n})^2,\\
&=k_1+\epsilon\Gamma_1\left[\frac{A}{2}\psi_0^be^{i\theta_n}+\frac{A}{2}(\psi_0^b)^*e^{-i\theta_n}-\frac{A}{2}\psi_0^ae^{i\theta_n}-\frac{A}{2}(\psi_0^a)^*e^{-i\theta_n}\right]^2
\\&=k_1+\frac{\epsilon\Gamma_1A^2}{4}\left\{2\left(\left\Vert\psi_0^b\right\Vert^2-\psi_0^b(\psi_0^a)^*-(\psi_0^b)^*\psi_0^a+\left\Vert\psi_0^a\right\Vert^2\right)+\left[(\psi_0^a)^2+(\psi_0^b)^2-2\psi_0^a\psi_0^b\right]e^{i2\theta_n}+c.c.\right\},
\end{split}
\end{align}
where $c.c.$ denotes the complex conjugate of the preceding
term. The first term on the right-hand side of Eq.\ref{EOMs_chains} can be written as 
\begin{align}\label{First_term_Eq4}
\begin{split}
k^e_1(u^b_{n}-u^a_{n})&=\frac{A}{2}\Bigg\{\left[k_1+\frac{\epsilon\Gamma_1A^2}{2}\left(\left\Vert\psi_0^b\right\Vert^2-\psi_0^b(\psi_0^a)^*-(\psi_0^b)^*\psi_0^a+\left\Vert\psi_0^a\right\Vert^2\right)\right](\psi^b-\psi^a)\\
 &+\frac{\epsilon\Gamma_1A^2}{4}\left[(\psi_0^a)^2+(\psi_0^b)^2-2\psi_0^a\psi_0^b\right]\left[(\psi^b)^*-(\psi^a)^*\right]\Bigg\} e^{i\theta_n}+\cdot\cdot\cdot\\
 &\approx \frac{A}{2}\Bigg\{\tilde{k}^e_1(\psi^b-\psi^a)+\frac{\epsilon\Gamma_1A^2}{4}\left[\left(2\left\Vert\psi_0^a\right\Vert^2+\left\Vert\psi_0^b\right\Vert^2-\psi_0^b(\psi_0^a)^*\right)\psi^{b}-\left(2\left\Vert\psi_0^b\right\Vert^2+\left\Vert\psi_0^a\right\Vert^2-\psi_0^a(\psi_0^b)^*\right)\psi^{a}\right]\Bigg\}e^{i\theta_n},
\end{split}
\end{align}
where $\tilde{k}^e_1=k_1+\frac{\epsilon\Gamma_1A^2}{2}\left(\left\Vert\psi_0^b\right\Vert^2-\psi_0^b(\psi_0^a)^*-(\psi_0^b)^*\psi_0^a+\left\Vert\psi_0^a\right\Vert^2\right)$.
\end{widetext}

To derive the equivalent linear eigenvalue problem, we only keep the term associated with $e^{i\theta_n}$, as shown in Eq.\ref{First_term_Eq4}. However, we note that the complex conjugates of the nonlinear eigenvector components $(\psi^{a,b})^*$ also appear in a component of the $e^{i\theta_n}$ term (the second line of Eq.\ref{First_term_Eq4}). To resolve this issue, we replace $(\psi^{a,b})^*$ by its linear counterpart $(\psi_0^{a,b})^*$ and return a linear eigenvector component $\psi_0^{a,b}$ back to the nonlinear one $\psi^{a,b}$. As a result, we get rid of all the $(\psi_0^{a,b})^*$ in the $e^{i\theta_n}$ term, as shown in the third line of Eq.\ref{First_term_Eq4}. Applying this treatment to other terms in Eq.\ref{EOMs_chains}, we can obtain the equivalent linear eigenvalue problem (Eq.\ref{EOMs_v1}) with the stiffness matrix $\bold{K}(\xi_0, A,\boldsymbol{\psi}_0)$.

 \subsection{Analytical solution of the equivalent linear eigenvalue problem}
Before solving Eq.~\ref{EOMs_v1}, we first obtain the normalized linear modal vector $\boldsymbol{\psi}_0=\begin{bmatrix}\psi_0^a\\ \psi_0^b\end{bmatrix}=\begin{bmatrix}1\\ \frac{-(k_1+k_2e^{i\xi_0})}{m\omega_0^2-k_1-k_2}\end{bmatrix}$. For the choice of the unit cell that contains two masses, there are two linear dispersion curves (and therefore two modal vectors) even in the linear regime where the system becomes a monatomic chain, as a result of the band folding effect. 

Using $\boldsymbol{\psi}_0$ as an initial guess, we can proceed to solve the equivalent linear eigenvalue problem (Eq.~\ref{EOMs_v1}), which leads to the following quadratic equation in terms of $\omega^2$:
\begin{equation}\label{Quadratic_eqn}
\omega^4+b\omega^2+c=0,
\end{equation}
where $b=\frac{K_{11}+K_{22}}{m}$ and $c=\frac{K_{11}K_{22}-K_{12}K_{21}}{m^2}$. 
Then, the nonlinear dispersion relation and modal vector can be analytically expressed as
\begin{equation}\label{nonlinear_dispersion}
\omega_{1,2}=\sqrt{\frac{b\mp \Delta}{2}} \,\,\, \text{and} \,\,\,
\boldsymbol{\psi}=\begin{bmatrix}1\\ \frac{-K_{21}}{m\omega_{1,2}^2-K_{22}}\end{bmatrix},
\end{equation}
where $\Delta=\sqrt{b^2-4c}$. For a given wavenumber $\xi_0$ and modal vector $\boldsymbol{\psi}_0$, we can obtain two nonlinear frequencies according to Eq.~\ref{nonlinear_dispersion}, one of which can be easily eliminated as it could obviously be infeasible. To obtain a more accurate result, we can set a tolerance and repeat the above procedures until the relative residual of the modal vector $\left\Vert\Delta \boldsymbol{\psi}\right\Vert$ satisfies the tolerance.

\subsection{Determination of the critical amplitude $A_c$}

\begin{figure}[htb]
    \centerline{ \includegraphics[width=0.45\textwidth]{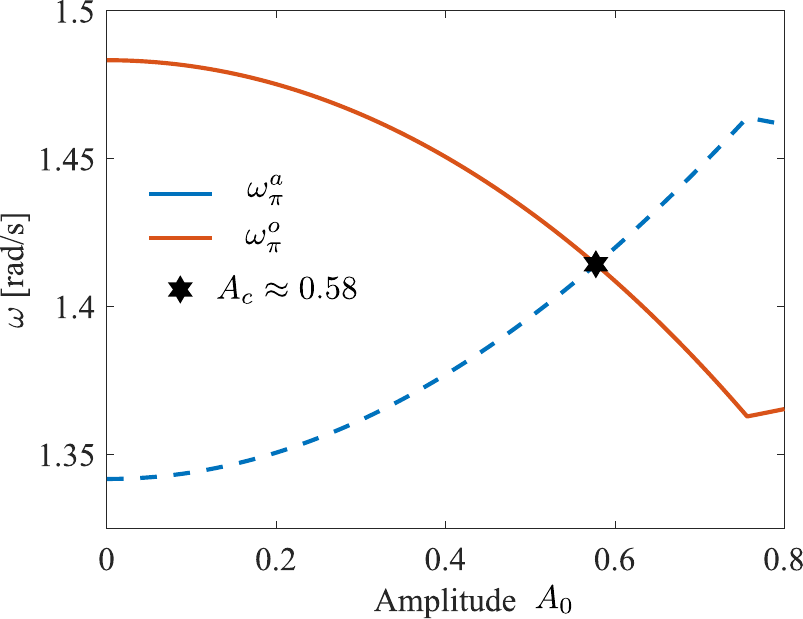}}
    \caption{Frequencies of the acoustic ($\omega_{\pi}^a$) and optical ($\omega_{\pi}^o$) branches at $\xi_0=\pi$ as functions of the excitation amplitude $A_0$. An intersection is observed between the two frequency curves, whose amplitude is identified as the critical amplitude $A_c\approx0.58$.} 
    \label{fig:Determination_Ac}
\end{figure}

To determine the critical amplitude $A_c$ at which the bandgap of the diatomic-like chain is just closed, we calculate for a wide range of excitation amplitude $A_0\in (0,0.8)$ the two frequencies of the acoustic and optical branches at $\xi_0=\pi$, denoted as $\omega_{\pi}^a$ and $\omega_{\pi}^o$, which correspond to the lower and upper bounds of the bandgap, respectively. As shown in Fig.~\ref{fig:Determination_Ac}, we identify an intersection (black star marker) between the two frequency curves, which indicates the onset of the bandgap closing phenomenon. Thus, the critical amplitude is determined as the corresponding amplitude at the intersection $A_c\approx0.58$.

\bibliographystyle{unsrt}
\bibliography{ref}
\clearpage
\onecolumngrid
\setcounter{figure}{0}
\setcounter{equation}{0}
\setcounter{page}{1}
\renewcommand{\thefigure}{S\arabic{figure}}
\renewcommand{\theequation}{S\arabic{equation}}
\section*{Supplemental information (SI)}
\subsection*{Note 1: Full expression of $\bold{K}(\xi_0, A,\boldsymbol{\psi}_0)$ for 1D nonlinear chains}
Based on the theoretical approach presented in Method, we derive the stiffness matrix $\bold{K}(\xi_0, A,\boldsymbol{\psi}_0)$ for the equivalent linear eigenvalue problem, whose components can be explicitly expressed as
\begin{align}\label{K_components}
\begin{split}
K_{11}&=-\tilde{k}^e_1-\tilde{k}^e_2+\frac{\epsilon A^2}{4}\left[\Gamma_1\left(\psi_0^a(\psi_0^b)^*-2\left\Vert\psi_0^b\right\Vert^2-\left\Vert\psi_0^a\right\Vert^2\right)+\Gamma_2\left(\psi_0^a(\psi_0^b)^*e^{i\xi}-\left\Vert\psi_0^a\right\Vert^2-2\left\Vert\psi_0^b\right\Vert^2\right)\right],\\
K_{12}&=\tilde{k}^e_1+\tilde{k}^e_2e^{-i\xi}+\frac{\epsilon A^2}{4}\left[\Gamma_1\left(2\left\Vert\psi_0^a\right\Vert^2+\left\Vert\psi_0^b\right\Vert^2-\psi_0^b(\psi_0^a)^*\right)+\Gamma_2\left(\left\Vert\psi_0^b\right\Vert^2e^{-i\xi}+2\left\Vert\psi_0^a\right\Vert^2e^{-i\xi}-\psi_0^b(\psi_0^a)^*e^{-i2\xi}\right)\right],\\
K_{21}&=(K_{12})^*,K_{22}=K_{11},
\end{split}
\end{align}
where 
\begin{align}\label{K_components_coe}
\begin{split}
\tilde{k}^e_1&=k_1+\frac{\epsilon\Gamma_1A^2}{2}\left(\left\Vert\psi_0^b\right\Vert^2-\psi_0^b(\psi_0^a)^*-(\psi_0^b)^*\psi_0^a+\left\Vert\psi_0^a\right\Vert^2\right),\\
\tilde{k}^e_2&=k_2+\frac{\epsilon\Gamma_2A^2}{2}\left(\left\Vert\psi_0^b\right\Vert^2-\psi_0^b(\psi_0^a)^*e^{-i\xi}-(\psi_0^b)^*\psi_0^ae^{i\xi}+\left\Vert\psi_0^a\right\Vert^2\right). 
\end{split}
\end{align}
Obviously, the stiffness matrix $\bold{K}$ is a Hermitian matrix.

\subsection*{Note 2: Multiple scales analysis for 1D nonlinear chains}
Here, we employ multiple scales analysis to derive nonlinear plane wave solutions of Eq.~\ref{EOMs_monatomic}. Based on previous work on 1D nonlinear spring-mass chains \cite{Jiao_pre_2019, Jiao_prsa_2021}, we introduce a fast spatiotemporal variable $\theta_{n,m}=\xi n -\omega t$ to capture the fundamental harmonic response, where $\xi$ and $\omega$ are the nondimensional wavenumber and frequency, respectively. To capture the weakly nonlinear effects (i.e., dispersion shifts), we introduce one slow spatial variable $s=\epsilon n$ and one slow temporal variable $\tau=\epsilon t$. The solution $\bold{u}_n=[u_{n}^a\,\,\,  u_{n}^b]^T$ is assumed to have an expansion of the form
\begin{equation}\label{solution_form}
\bold{u}_{n}=\bold{u}^0_{n}(\theta_{n},s,\tau)+\epsilon\bold{u}^1_{n}(\theta_{n},s,\tau)+\bold{O}(\epsilon^2).
\end{equation}
According to Bloch's theorem, the relation between displacements at neighboring unit cells reads 
\begin{equation}\label{Bloch_conds}
\bold{u}_{n\pm1}=\bold{u}_{n}e^{\pm i\xi}.
\end{equation}

Substituting Eqs.~\ref{solution_form}-\ref{Bloch_conds} in Eq.~1 yields the following cascading equation up to the order of $O(\epsilon)$:
\begin{align}
O(1): \bold{M}\ddot{\bold{u}}^0_{n}+\bold{K}\bold{u}^0_{n}&=\bold{0},\label{EOMs_cascading_0}\\
O(\epsilon): \bold{M}\ddot{\bold{u}}^1_{n}+\bold{K}\bold{u}^1_{n}&=\bold{f},\label{EOMs_cascading_1}
\end{align}
where $\bold{M}=\begin{bmatrix} m & 0\\ 0 & m \end{bmatrix}$ is a mass matrix, $\bold{K}=\begin{bmatrix} k_1+k_2 & -k_1-k_2e^{-i\xi}\\ -k_1-k_2e^{i\xi} & k_1+k_2 \end{bmatrix}$ is the stiffness matrix for the linear system, and $\bold{f}$ is a forcing term at the order of $O(\epsilon)$. The expression of $\bold{f}$ is given as 
\begin{equation}\label{forcing_term}
\bold{f}=2\omega\bold{M}\frac{\partial^2 \bold{u}_n^0}{\partial \theta_n \partial \tau}+\begin{bmatrix} -k_2\frac{\partial  (u^b_{n-1})^0}{\partial s} +\Gamma_1\left[(u^b_n)^0-(u^a_n)^0\right]^3+\Gamma_2\left[(u^b_{n-1})^0-(u^a_n)^0\right]^3\\  k_1\frac{\partial  (u^a_{n+1})^0}{\partial s}+\Gamma_1\left[(u^a_n)^0-(u^b_n)^0\right]^3+\Gamma_2\left[(u^a_{n+1})^0-(u^b_n)^0\right]^3 \end{bmatrix}.
\end{equation}

Eq.~\ref{EOMs_cascading_0} is the linearized equation of motion for the nonlinear chain, whose plane wave solution can be written as 
\begin{equation}\label{order1_soln}
\bold{u}^0_{n}=\frac{A(s,\tau)}{2}\boldsymbol{\psi}_0e^{i\theta_{n}}+\frac{A^*(s,\tau)}{2}\boldsymbol{\psi}_0^*e^{-i\theta_{n}},
\end{equation}
where $A(s,\tau)$ is an arbitrary function of the slow variables $(s,\tau)$, and $\boldsymbol{\psi}=[\psi_0^a\,\,\,  \psi_0^b]^T$ is the linear modal vector. It should be noted that $(\xi,\omega)$ denotes the wavenumer-frequency pair of the linear dispersion relation.

At order $O(\epsilon)$, Eq.~\ref{EOMs_cascading_1} has the same kernel as Eq.~\ref{EOMs_cascading_0} albeit with a forcing term $\bold{f}$. After substituting Eq.~\ref{order1_soln} in Eq.~\ref{forcing_term} and premultiply by the transpose conjugate of $\boldsymbol{\psi}_0$ (i.e., $\boldsymbol{\psi}_0^H$), the terms associated with $e^{\pm i\theta_{n}}$ are identified as secular terms in $\bold{f}$, which must be eliminated to avoid any unbounded solution. This treatment leads to the following condition:
\begin{equation}\label{Scular_terms}
\frac{\partial A}{\partial \tau}+\mu\frac{\partial A}{\partial s}=-i\lambda\left\Vert A\right\Vert^2,
\end{equation}
where $\mu=\left[-e^{-i\xi}(\psi_0^a)^*\psi_0^b+e^{i\xi}\psi_0^a(\psi_0^b)^*\right]/(2\omega \Bar{m})$ and $\lambda=-3\gamma/(8\omega \Bar{m})$, in which $\Bar{m}$ and $\gamma$ can be expressed as
\begin{equation}\label{m_bar}
\begin{split}
\Bar{m}&=\boldsymbol{\psi}_0^H\bold{M}\boldsymbol{\psi}_0=\left(\left\Vert\psi_0^a\right\Vert^2+\left\Vert\psi_0^b\right\Vert^2\right)m,
\end{split}
\end{equation}
and
\begin{align}\label{gamma}
\begin{split}
\gamma&=\Gamma_1(\psi_0^a)^*\left(\left\Vert\psi_0^b\right\Vert^2\psi_0^b-(\psi_0^b)^2(\psi_0^a)^* +(\psi_0^a)^2(\psi_0^b)^*-2\left\Vert\psi_0^b\right\Vert^2\psi_0^a+2\left\Vert\psi_0^a\right\Vert^2\psi_0^b-\left\Vert\psi_0^a\right\Vert^2\psi_0^a\right)\\
&+\Gamma_2(\psi_0^a)^*\left(\left\Vert\psi_0^b\right\Vert^2\psi_0^be^{-i\xi}-(\psi_0^b)^2(\psi_0^a)^*e^{-i2\xi} +(\psi_0^a)^2(\psi_0^b)^*e^{i\xi}-2\left\Vert\psi_0^b\right\Vert^2\psi_0^a+2\left\Vert\psi_0^a\right\Vert^2\psi_0^be^{-i\xi}-\left\Vert\psi_0^a\right\Vert^2\psi_0^a\right)\\
&+\Gamma_2(\psi_0^b)^*\left(\left\Vert\psi_0^a\right\Vert^2\psi_0^ae^{i\xi}-(\psi_0^a)^2(\psi_0^b)^*e^{i2\xi} +(\psi_0^b)^2(\psi_0^a)^*e^{-i\xi}-2\left\Vert\psi_0^a\right\Vert^2\psi_0^b+2\left\Vert\psi_0^b\right\Vert^2\psi_0^ae^{i\xi}-\left\Vert\psi_0^b\right\Vert^2\psi_0^b\right)\\
&-\Gamma_1(\psi_0^b)^*\left(\left\Vert\psi_0^b\right\Vert^2\psi_0^b-(\psi_0^b)^2(\psi_0^a)^* +(\psi_0^a)^2(\psi_0^b)^*-2\left\Vert\psi_0^b\right\Vert^2\psi_0^a+2\left\Vert\psi_0^a\right\Vert^2\psi_0^b-\left\Vert\psi_0^a\right\Vert^2\psi_0^a\right).
\end{split}
\end{align}

To simplify the derivation, we write $A$ in a polar form as
\begin{equation}\label{A_polar}
A=\alpha(s,\tau)e^{-i\beta(s,\tau)},
\end{equation}
where $\alpha$ and $\beta$ are real numbers and functions of $(s_1,s_2,\tau)$. Substituting Eq.~\ref{A_polar} in Eq.~\ref{Scular_terms} and separating the real and imaginary components yields the following equations:
\begin{align}\label{condition_polar}
\begin{split}
\frac{\partial \alpha}{\partial \tau}+\mu\frac{\partial \alpha}{\partial s}&=0,\\
\frac{\partial \beta}{\partial \tau}+\mu\frac{\partial \beta}{\partial s}&=\lambda\alpha^2.
\end{split}
\end{align}
The general solution of Eq.~\ref{condition_polar} are 
\begin{align}\label{condition_soln}
\begin{split}
\alpha&=\alpha_0(s-\mu\tau),\\
\beta&=\beta_0(s-\mu\tau)+\beta^*,
\end{split}
\end{align}
where $\alpha_0$ and $\beta_0$ are functions of $s-\mu\tau$, and $\beta^*$ is a special solution of $\beta$. Interestingly, $\beta^*$ can be expressed either in terms of spatial variables $s$ as $\beta^*=\lambda\alpha^2s_1/\mu$, or in terms of temporal variable $\tau$ as $\beta^*=-\lambda\alpha^2\tau$. The specific form of $\alpha$, $\beta$, and $\beta^*$ can be determined using excitation conditions. For an initial condition (e.g., spatial profile $\bold{u}_{n} =A_0\boldsymbol{\psi}_0 \cos (\xi n)$ at $t=0$), it follows that $\alpha_0=A_0$, $\beta_0=0$, and $\beta^*=\lambda A_0^2\tau$. Combined with Eq.~\ref{order1_soln} and Eq.~\ref{A_polar}, the fundamental solution at $O(1)$ is obtained as
\begin{equation}\label{Soln_initial}
\bold{u}^0_{n}=\frac{A_0}{2}\boldsymbol{\psi}_0e^{i\left[\xi n -(\omega +\epsilon\lambda A_0^2)t\right]}+c.c.,
\end{equation}
where $c.c.$ denotes the complex conjugate of the preceding term. Thus, the nonlinear effect manifests as frequency correction under initial excitation conditions. For the nonlinear chain with alternating softening and hardening nonlinearities, the nonlinear dispersion relation predicted from Eq.~\ref{Soln_initial} is given in Fig.~1 (diamond markers), which is identical to the linear one without bandgap opening. The failure of the above multiple scales analysis to capture the bandgap opening phenomenon is because of the 
inability to resolve the nonlinear modal vector, as explained in the main text. 

\subsection*{Note 3: Full expression of $\bold{K}(\boldsymbol{\kappa}, A,\boldsymbol{\psi}_{i-1})$ for 2D nonlinear hexagonal lattices}
Applying the theoretical approach to 2D nonlinear hexagonal lattices, we derive the stiffness $\bold{K}(\boldsymbol{\kappa}, A,\boldsymbol{\psi}_{i-1})$ for the equivalent linear eigenvalue problem. The components of $\bold{K}$ can be explicitly expressed as
\begin{align}\label{K_components_hexagonal}
\begin{split}
K_{11}&=-\tilde{k}^e_1-\tilde{k}^e_2-\tilde{k}^e_3+\frac{\epsilon A^2}{4}\Bigg\{\Gamma_1\left(\psi_{i-1}^a(\psi_{i-1}^b)^*-2\left\Vert\psi_{i-1}^b\right\Vert^2-\left\Vert\psi_{i-1}^a\right\Vert^2\right)\\&+\Gamma_2\left[\psi_{i-1}^a(\psi_{i-1}^b)^*(e^{i\xi_1}+e^{i\xi_2})-2\left\Vert\psi_{i-1}^a\right\Vert^2-4\left\Vert\psi_{i-1}^b\right\Vert^2\right]\Bigg\},\\
K_{12}&=\tilde{k}^e_1+\tilde{k}^e_2e^{-i\xi_1}+\tilde{k}^e_3e^{-i\xi_2}+\frac{\epsilon A^2}{4}\Bigg\{\Gamma_1\left(2\left\Vert\psi_{i-1}^a\right\Vert^2+\left\Vert\psi_{i-1}^b\right\Vert^2-\psi_{i-1}^b(\psi_{i-1}^a)^*\right)\\&+\Gamma_2\left[\left\Vert\psi_{i-1}^b\right\Vert^2(e^{-i\xi_1}+e^{-i\xi_2})+2\left\Vert\psi_{i-1}^a\right\Vert^2(e^{-i\xi_1}+e^{-i\xi_2})-\psi_{i-1}^b(\psi_{i-1}^a)^*(e^{-i2\xi_1}+e^{-i2\xi_2})\right]\Bigg\},\\
K_{21}&=(K_{12})^*,K_{22}=K_{11},
\end{split}
\end{align}
where $\xi_1$ and $\xi_2$ are the nondimensional wavenumbers in the reciprocal lattice vector basis, and 
\begin{align}\label{K_components_hexagonal}
\begin{split}
\tilde{k}^e_1&=k_1+\frac{\epsilon\Gamma_1A^2}{2}\left(\left\Vert\psi_{i-1}^b\right\Vert^2-\psi_{i-1}^b(\psi_{i-1}^a)^*-(\psi_{i-1}^b)^*\psi_{i-1}^a+\left\Vert\psi_{i-1}^a\right\Vert^2\right), \\\tilde{k}^e_2&=k_2+\frac{\epsilon\Gamma_2A^2}{2}\left(\left\Vert\psi_{i-1}^b\right\Vert^2-\psi_{i-1}^b(\psi_{i-1}^a)^*e^{-i\xi_1}-(\psi_{i-1}^b)^*\psi_{i-1}^ae^{i\xi_1}+\left\Vert\psi_{i-1}^a\right\Vert^2\right),\\
\tilde{k}^e_3&=k_2+\frac{\epsilon\Gamma_2A^2}{2}\left(\left\Vert\psi_{i-1}^b\right\Vert^2-\psi_{i-1}^b(\psi_{i-1}^a)^*e^{-i\xi_2}-(\psi_{i-1}^b)^*\psi_{i-1}^ae^{i\xi_2}+\left\Vert\psi_{i-1}^a\right\Vert^2\right).
\end{split}
\end{align}
Obviously, the stiffness matrix $\bold{K}$ is a Hermitian matrix.
\end{document}